\documentclass[aps,prb,twocolumn,showpacs,floatfix]{revtex4}

\usepackage{graphicx}
\usepackage{dcolumn}
\usepackage{bm}
\usepackage{times}
\usepackage{color}
\begin{document}
\bibliographystyle{apsrev}
\title{Magnetoresistance and Shubnikov-de Hass oscillation in YSb}
\author{Qiao-He Yu}\thanks{These authors contributed equally to this paper}
\author{Yi-Yan Wang}\thanks{These authors contributed equally to this paper}
\author{Rui Lou}
\author{Peng-Jie Guo}
\author{Sheng Xu}
\author{Kai Liu}
\author{Shancai Wang}
\author{Tian-Long Xia}\email{tlxia@ruc.edu.cn}
\affiliation{Department of Physics, Renmin University of China, Beijing 100872, P. R. China}
\affiliation{Beijing Key Laboratory of Opto-electronic Functional Materials $\&$ Micro-nano Devices, Renmin University of China, Beijing 100872, P. R. China}

\date{\today}
\begin{abstract}
YSb crystals are grown and the transport properties under magnetic field are measured. The resistivity exhibits metallic behavior under zero magnetic field and the low temperature resistivity shows a clear upturn once a moderate magnetic field is applied. The upturn is greatly enhanced by increasing magnetic field, finally resulting in a metal-to-insulator-like transition. With temperature further decreased, a resistivity plateau emerges after the insulator-like regime. At low temperature (2.5 K) and high field (14 T), the transverse magnetoresistance (MR) is quite large ($3.47\times10^4\%$). In addition, Shubnikov-de Haas (SdH) oscillation has also been observed in YSb. Periodic behavior of the oscillation amplitude reveals the related information about Fermi surface and two major oscillation frequencies can be obtained from the FFT spectra of the oscillations. The trivial Berry phase extracted from SdH oscillation, band structure revealed by angle-resolved photoemission spectroscopy (ARPES) and first-principles calculations demonstrate that YSb is a topologically trivial material.
\end{abstract}
\maketitle
\setlength{\parindent}{1em}
\section{Introduction}

The magnetoresistance (MR) effect, which describes the change of resistance induced by the magnetic field, is an attractive topic in condensed matter physics. MR not only has led to many important applications such as magnetic field sensors, but also is a useful way to obtain information about electronic structure of conductors\cite{shoenberg1984magnetic,pippard1989magnetoresistance}. In the past several decades, the in-depth study of giant magnetoresistance (GMR) in magnetic multilayers\cite{baibich1988giant,binasch1989enhanced} and colossal magnetoresistance (CMR) in magnetic oxide materials\cite{moritomo1996giant,ramirez1997colossal} has broadened people's understanding of MR in materials greatly. Recently,  much attention has been paid on the extremely large MR (XMR) around 10$^5$\% to 10$^6$\%. The XMR has been detected in several nonmagnetic materials, such as TX (T=Ta/Nb, X=As/P)\cite{weng2015weyl,huang2015observation,ghimire2015magnetotransport,yang2015chiral,shekhar2015large,hu2016pi,zhang2015large,shekhar2015extremely,wang2015helicity}, TX$_2$ (T=Ta/Nb, X=As/Sb)\cite{wang2014anisotropic,li2016field,wang2016resistivity,wu2016giant,luo2016anomalous,yuan2016large,shen2016fermi,xu2016electronic,wang2016topological,li2016negative}, LaX(X=Sb/Bi)\cite{tafti2015resistivity,tafti2016temperature,sun2016large,kumar2016observation,guo2016perfect}, Cd$_3$As$_2$\cite{liang2015ultrahigh,he2014quantum}, and WTe$_2$\cite{ali2014large,PhysRevLett.115.046602,wang2015origin,ali2015correlation,zhu2015quantum,PhysRevLett.115.166602,PhysRevB.92.125152} etc.

Recently, the rare earth-based materials LaSb and LaBi with simple rock salt structure have trigged great interest. Both of them exhibit XMR and field-induced metal-to-insulator-like transition followed by a resistivity plateau at low temperature and high magnetic field. The transverse MR even reaches $9\times 10^5\%$ for LaSb at 2 K and 9 T. SdH oscillation and high mobility have also been observed. However, the origin of the XMR and field-induced transport properties is still controversial. One view attributes it to the compensation of hole and electron, where the electron-hole balance and high mobility result in the quadratic behavior and very large value of MR\cite{sun2016large,kumar2016observation,guo2016perfect}. Another view thinks that XMR is possibly the consequence of a combination of electron-hole compensation and the mixed $d$-$p$ orbital texture by a magnetic field\cite{tafti2016temperature}. To reveal the physics underneath, it is in great demand to find more materials with similar properties.

In this paper, we report the growth of YSb single crystals with the same crystal structure and similar chemical composition to LaSb/LaBi. The magneto-transport properties of YSb have been studied in detail. YSb exhibits similar field-induced behavior and XMR as in LaSb/LaBi. The two major frequencies obtained from FFT spectra of SdH quantum oscillation are larger than those in LaSb/LaBi, which reveals a larger cross sectional area of Fermi surface in YSb. The trivial Berry phase, ARPES results and band structure calculation all demonstrate that YSb is a topologically trivial material. The electron-hole compensation is suggested to be responsible for the XMR in YSb.

\begin{figure}[htbp]
\centering
\includegraphics[width=0.48\textwidth]{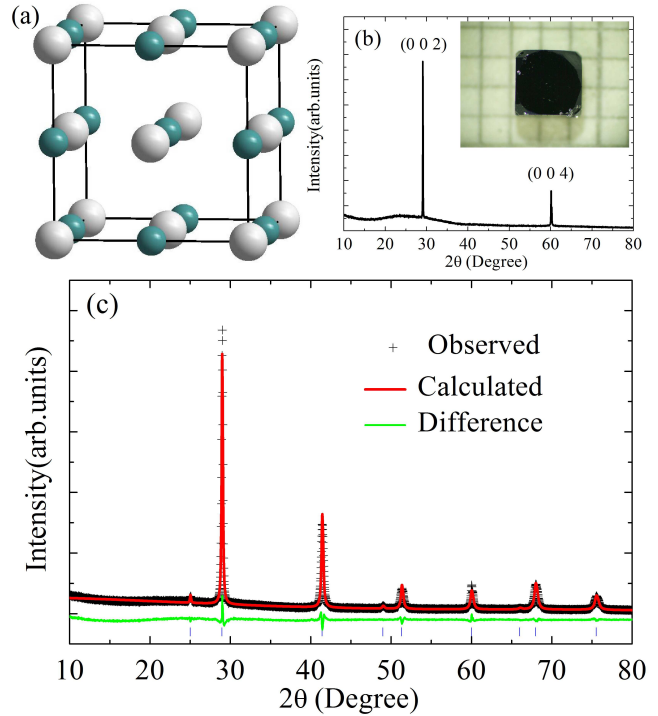}
\caption{(Color online) (a) Crystal structure of YSb. The white and teal balls represent Y and Sb, respectively. (b) XRD of a single crystal. Inset: the photo of a selected single crystal (each grid stands for 1$\times$1mm$^2$). (c) Refinement results of powder XRD using TOPAS, a=6.1628(6)$\AA$ and R$_{wp}$=7.914\%.}
\end{figure}

\section{Methods and crystal structure}

Single crystals of YSb were grown with Antimony flux method. The starting elements of Y(99.6\%) and Sb(99.5\%) were placed into an alumina crucible and sealed in a quartz tube. The quartz tube was put in a high temperature furnace and heated to 1273 K, held for several hours, then cooled down to 1023 K within 200 hours. At this temperature, the excess Sb flux was removed with centrifuge. The atomic proportion confirmed by energy dispersive x-ray spectroscopy (EDX) was consistent with 1:1 for Y:Sb. X-ray diffraction (XRD) patterns of single crystal and crushed crystal powder were obtained using a Bruker D8 Advance x-ray diffractometer. TOPAS-4.2 was employed for the refinement. Resistivity measurements were performed with four-probe method in physical property measurement system (Quantum Design PPMS-14T). ARPES measurements were performed at the Dreamline beam line of the Shanghai Synchrotron Radiation Facility (SSRF) with a Scienta D80 analyzer. The energy and angular resolutions were set to 15 meV and 0.05бу, respectively. The samples were cleaved in situ along the (0 0 1) plane and measured at T=30 K in a working vacuum better than 5$\times$10$^{-11}$ Torr. The electronic structures of YSb have been studied by using the first-principles calculations. The projector augmented wave (PAW) method\cite{PhysRevB.50.17953,PhysRevB.59.1758} as implemented in the VASP package\cite{PhysRevB.47.558,kresse1996efficiency,PhysRevB.54.11169} was used to describe the core electrons. For the exchange-correlation potential, the modified Becke-Johnson (MBJ)\cite{becke2006simple,PhysRevLett.102.226401} exchange potential with the GGA correlation was used. The kinetic energy cutoff of the plane-wave basis was set to be 300 eV. A $20\times20\times20$ k-point mesh was utilized for the Brillouin zone (BZ) sampling and the Fermi surface was broadened by the Gaussian smearing method with a width of 0.05 eV. Both cell parameters and internal atomic positions were allowed to relax until all forces were smaller than 0.01eV/${\AA}$. The calculated equilibrium lattice constant a is 6.202${\AA}$, which agrees well with the previous experimental value 6.163${\AA}$ \cite{Brixner1960Structure} and the value obtained with x-ray refinement in this paper. Once the equilibrium crystal structures were obtained, the electronic structures were calculated by including the spin orbital coupling (SOC) effect. The Fermi surfaces were studied by using the maximally localized Wannier functions (MLWF)\cite{PhysRevB.56.12847,PhysRevB.65.035109}.

YSb crystallizes in a rock salt structure as shown in Fig. 1(a). The XRD pattern of a selected crystal shown in Fig. 1(b) indicates that the surface of the crystal is the (0 0 1) plane. The powder XRD pattern of YSb crystal is shown in Fig. 1(c). It was refined using the face-centered cubic structure with space group Fm-3m (No.225) and the refined lattice parameter a is 6.1628(6)$\AA$, which is in good agreement with the value found from the Inorganic Crystal Structure Database (ICSD). As shown in the photo of Fig. 1(b), the typical size of YSb crystals is about $2\times2\times2$$mm^3$.

\begin{figure}[htbp]
\centering
\includegraphics[width=0.48\textwidth]{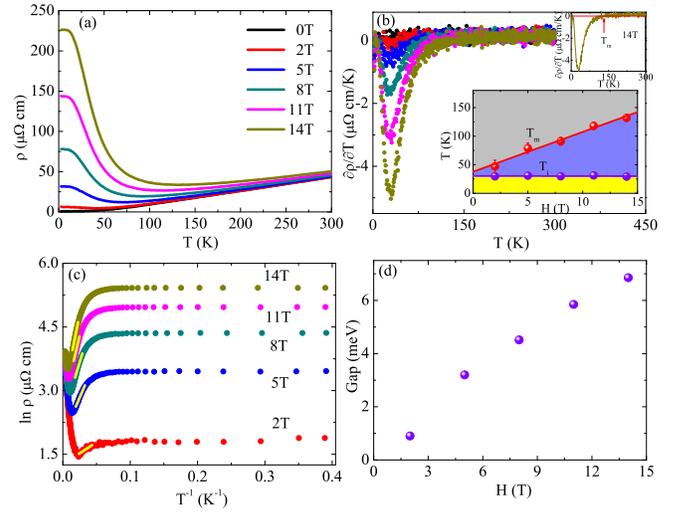}
\caption{(Color online) (a) Resistivity as a function of temperature
in YSb with several magnetic fields applied. (b) $\partial
\rho/\partial T$ as a function of temperature. Upper inset:
$\partial \rho/\partial T$ as a function of temperature under 14 T.
The red arrow denotes the T$_m$. Lower inset: Temperature of
resistivity minimum T$_m$ and inflection T$_i$ as a function of
magnetic field. (c) Plot of $ln(\rho)$ as a function of T$^{-1}$
used to achieve the energy gap at several fields. (d) The extracted
energy gap values from Fig. 2(c) as a function of magnetic field. }
\end{figure}

\begin{figure}[htbp]
\centering
\includegraphics[width=0.48\textwidth]{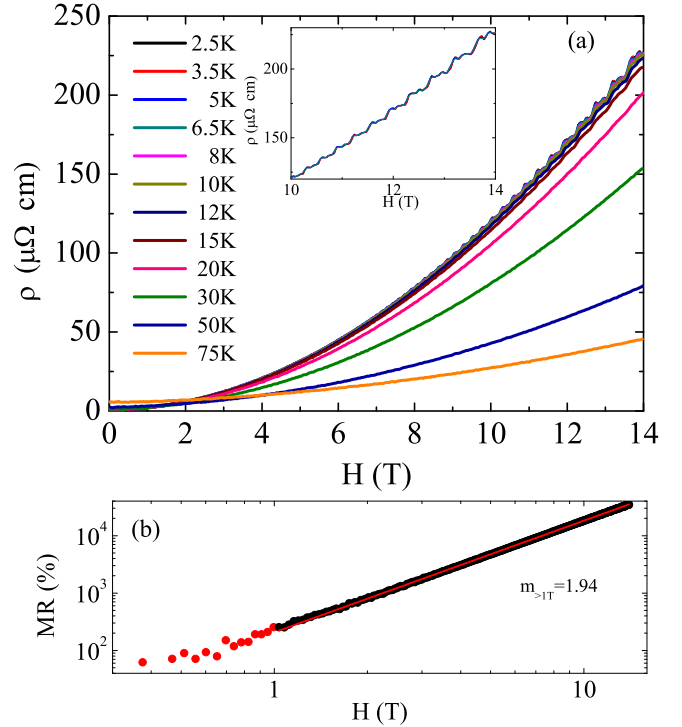}
\caption{(Color online) (a) Resistivity as a function of magnetic field in YSb with RRR=63 at several temperatures. Inset: enlarged parts of resistivity at low temperatures (2.5 K, 3.5 K, 5 K, 6.5 K) under high fields. (b) The logarithmic plot of the MR-H curve at 2.5 K. }
\end{figure}

\section{Results and discussions}

Figure 2(a) plots the temperature dependence of resistivity under
several magnetic fields. The electric current is parallel to (0 0 1)
plane and the magnetic field is parallel to [0 0 1] direction. The
temperature dependent resistivity at zero field exhibits a metallic
behavior. The high residual resistivity ratio (RRR,
$\rho_{300K}/\rho_{2.5K}=63$) indicates good quality of the samples.
When the magnetic field is applied, resistivity decreases with
decreasing temperature until a minimum T$_m$, then increases until
an inflection at T$_i$ where the resistivity plateau starts to
emerge. A metal-to-insulator-like transition is observed at low
temperature when a moderate magnetic field is applied. Such behavior
is also observed in previous studies where the mechanism was under
debate\cite{PhysRevLett.94.166601,PhysRevLett.90.156402,PhysRevLett.87.206401,wang2015origin,PhysRevLett.115.046602}.
YSb exhibits large transverse MR ($[\rho_{xx}(H)-\rho_{xx}(0
T)]/\rho_{xx}(0 T)\times100\%$) of $3.47\times10^4\%$ at 2.5 K under
the field of 14 T. According to the previous study, MR will increase
with increasing
RRR\cite{tafti2016temperature,ali2015correlation,tafti2015resistivity}.
As a result, we could get a larger MR for YSb if we can improve the
quality of single crystals. Figure 2(b) shows the $\partial
\rho/\partial T$ curves derived from Fig. 2(a). T$_m$ is defined as
the temperature where the sign changes which can be seen clearly in
the upper inset and T$_i$ is defined as the temperature where a
valley appears. The inset shows T$_m$ and T$_i$ as a function of
field, and it is clearly shown that T$_i$ nearly keeps constant and
T$_m$ nearly increases linearly with increasing field. Figure 2(c)
plots the $ln(\rho)$ as a function of the reciprocal of temperature.
The values of energy gap can be obtained by fitting the
insulator-like regions (the linear part in Fig. 2(c)) using the
relation $\rho(T) \propto exp(E_a/k_BT)$ where $E_a$ is the energy
gap and $k_B$ is the Boltzmann constant. As shown in Fig. 2(d), the
energy gap increases with increasing field, which resembles previous
reports in LaSb and
LaBi\cite{tafti2015resistivity,tafti2016temperature}.

\begin{figure}[htbp]
\centering
\includegraphics[width=0.48\textwidth]{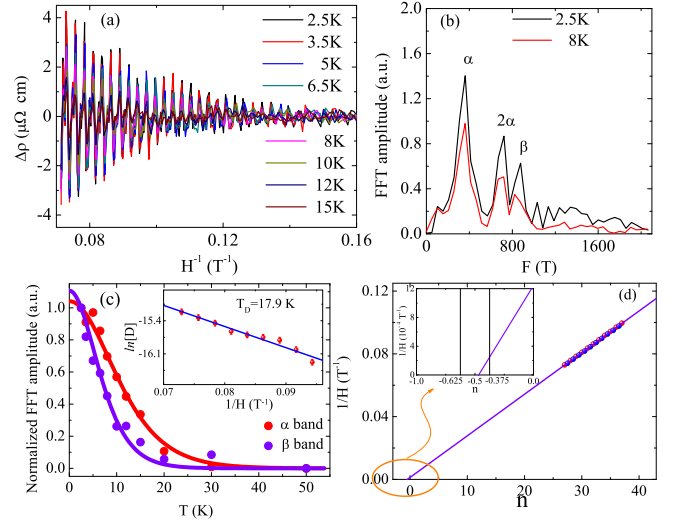}
\caption{(Color online) (a) The amplitude of SdH oscillations as a function of H$^{-1}$ at several temperatures. (b) Fast Fourier Transform of the corresponding SdH oscillations at 2.5 K and 8 K. (c) Normalized amplitude of the fast Fourier transform of SdH oscillations with two different oscillation frequencies plotted as a function of temperature for YSb. Solid lines are fitting curves using the Lifshitz-Kosevitch formula. The effective masses of carriers are extracted from the fits. Inset: $ln[D]$ vs $1/H$ at T=2.5 K. The solid line is linear fit to obtain the Dingle temperature. (d) 1/H as a function of the Landau level indices \emph{n} for the $\alpha$ band. The red and blue symbols correspond to the positions of peaks and valleys in $\Delta \rho$ curve at 2.5 K. The violet curve stands for the linear fit of data. Inset: local enlarged drawing of Fig. 4(d). }
\end{figure}

Figure 3(a) shows the resistivity of YSb as a function of field at different temperatures. Clear Shubnikov-de Haas (SdH) oscillation was observed at low temperature and high field. The inset shows the enlarged images of oscillating parts. With the increase of temperature, the transverse MR becomes smaller and the oscillation gradually disappears. The logarithmic plot of the MR-H curve at 2.5 K is shown in Fig. 3(b). It shows that the MR follows a semiclassical non-saturating quadratic behavior ($MR\sim H^m$ with $m\approx2$). According to the semiclassical two-band model, MR exhibits quadratic behavior at low field and reaches saturation at high field. However, in the case of electron-hole compensation, that is $n_e=n_h$ ($n_e$ and $n_h$ correspond to electron concentration and hole concentration), the model gives $MR=\mu_e\mu_h H^2$ ($\mu_e$ and $\mu_h$ correspond to electron mobility and hole mobility). That means the MR will follow a non-saturating quadratic behavior. The field dependent MR suggests that YSb is possibly a electron-hole compensated semimetal.

The oscillation part of resistivity is obtained by subtracting a smooth background. Figure 4(a) plots the oscillation amplitude $\Delta \rho_{xx}= \rho_{xx}-\langle \rho_{xx} \rangle$ of YSb against the reciprocal of magnetic field at various temperatures. The amplitude displays an obvious periodic behavior and decreases with increasing temperature or decreasing field. The oscillation amplitude can be described by the Lifshitz-Kosevich (L-K) formula\cite{shoenberg1984magnetic},
\begin{equation}\label{equ1}
\Delta\rho_{xx}\propto\frac{\lambda T}{sinh(\lambda T)}e^{-\lambda T_D}cos[2\pi\times(\frac{F}{H}-\frac{1}{2}+\beta+\delta)]
\end{equation}
where $\lambda= (2\pi^2k_{B}m^*)/(\hbar e\bar{H})$, $m^*$ is the effective mass of carrier, and $k_{B}$ is Boltzmann's constant. $T_D$ and $2\pi \beta$ are the Dingle temperature and Berry phase, respectively. $\delta$ is a phase shift with the value of $\delta=0$ (or $\pm1/8$) for 2D (or 3D) system\cite{luk2004phase}. Figure 4(b) shows the fast Fourier transformation (FFT) spectra at 2.5 K and 8 K. It shows that there exist two principle oscillation frequencies, $F_{\alpha}$=361 T with its second harmonic frequency $F_{2\alpha}$=722 T and $F_{\beta}$=877 T. In SdH oscillation, the frequency $F$ is proportional to the cross sectional area $A$ of Fermi surface normal to the magnetic field, which can be described using Onsager relation $F= (\phi_0/2\pi^2)A=(\hbar/2\pi e)A$. These frequencies in YSb are higher than those in LaSb and LaBi\cite{wang2015origin,tafti2016temperature,tafti2015resistivity,kumar2016observation,sun2016large}, revealing YSb has larger Fermi surfaces than LaSb and LaBi. In Fig. 4(c), we display the temperature dependence of the relative FFT amplitude of frequencies $\alpha$ and $\beta$ of YSb, respectively. The thermal factor $R_T=(\lambda T)/sinh(\lambda T)$ in L-K formula has been employed to describe the temperature dependence of FFT amplitude.
The effective masses $m^*_\alpha=0.17m_e$ and $m^*_\beta=0.27m_e$ can be extracted from the fits. We analyze the main frequency appearing at 361 T to obtain some basic parameters related to the Fermi surface. Considering the circular cross section of the Fermi surface along [0 0 1], the area can be obtained to be $3.45\times10^{-2}$ ${\AA}^{-2}$ by using Onsager relation. This value is only 3.3\% of the whole Brillouin zone in $k_x - k_y$ plane taking account of the lattice parameter $a=6.1628{\AA}$. The Fermi vector is found out to be 0.105${\AA}^{-1}$  by using the value of cross sectional area. The values of the Fermi velocity and the Fermi energy are $7.2\times10^5 m/s$ and 0.502 eV due to the relations $v_F=\hbar k_F/m^*$ and $m^*=E_F/v_F^2$, respectively. The Dingle temperature $T_D$=17.9 K is obtained from the slope in the plot of $ln[D]=\Delta\rho_{xx}Hsinh(\lambda T)$ versus $1/H$ at T=2.5 K as shown in the inset of Fig. 4(c). The corresponding quantum lifetime is $\tau_Q^\alpha=\hbar/2\pi k_B T_D=6.8\times10^{-14} s$ . Figure 4(d) shows the 1/H curve as a function of the Landau level indices \emph{n} for the $\alpha$ band. The Berry phase can be extracted based on Lifshitz-Onsager quantization rule $F/H=n+1/2-\beta+\delta$. The peaks and valleys of the $\Delta \rho$ at 2.5 K are denoted as integer and half-integer Landau level indices, respectively. The linear fitting gives a result of $1/2-\beta+\delta=0.4679$, indicating a trivial Berry phase for the $\alpha$ band.

\begin{figure}[htbp]
\centering
\includegraphics[width=0.48\textwidth]{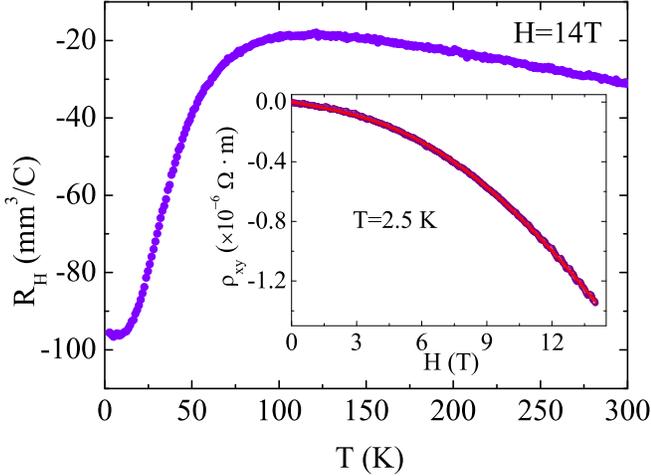}
\caption{(Color online) Hall coefficient at H=14 T plotted as a function of temperature from T=2.5 K to 300 K in YSb. Inset: Hall resistivity plotted as a function of field from H=0 T to 14 T at 2.5 K. The red solid line is the fit using two-band model.}
\end{figure}

Figure 5 shows temperature dependence of the Hall coefficient $R_H=\rho_{xy}/H$ at 14 T in YSb. $R_H$ in YSb increases sharply during low temperatures and then decreases slowly at high temperatures, which is similar to LaSb. However, YSb shows negative $R_H$ up to 300 K, while LaSb shows negative $R_H$ below 40 K and positive above 40 K that undergoes a second sign change at 170 K \cite{tafti2016temperature}. The field dependent Hall resistivity at 2.5 K is shown in the inset of Fig. 5. It shows clear oscillation, which also indicates the Landau level emptying as the magnetic field increases. The nonlinear behavior of $\rho_{xy}$ indicates the coexistence of electron and hole in YSb. According to the two-band model,
\begin{equation}\label{equ2}
\rho_{xy}=\frac{H}{e}\frac{(n_h \mu_h^2-n_e \mu_e^2)+(n_h-n_e)(\mu_h \mu_e)^2 H^2}{(n_h \mu_h+n_e \mu_e)^2+(n_h-n_e)^2 (\mu_h \mu_e)^2 H^2}
\end{equation}
the curve (the red solid line) is consistent with the experimental curve (the black solid line). At 2.5 K, the concentrations and mobilities obtained from the two-band model are $n_e=2.06\times10^{20}cm^{-3}$,$n_h=1.74\times10^{20}cm^{-3}$,$\mu_e=4.02\times10^4cm^2V^{-1}s^{-1}$, and $\mu_h=0.42\times10^4cm^2V^{-1}s^{-1}$. The similar concentrations suggests the compensation of electron and hole in YSb. However, the ratio $n_h/n_e\approx0.84$ indicates the compensation is not perfect.

\begin{figure}[htbp]
\centering
\includegraphics[width=0.48\textwidth]{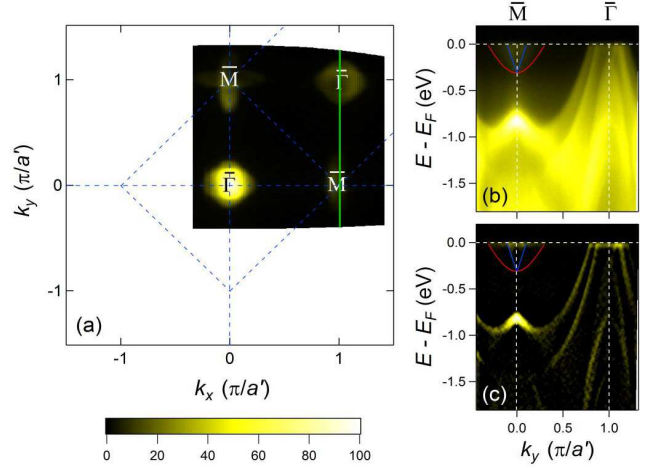}
\caption{(Color online) (a) ARPES intensity plot of YSb at $E_{F}$ as a function of the 2D wave vector recorded with $h\nu$=53 eV at T=30 K. The intensity plot is obtained by integrating the spectra within $E_{F}\pm$10 meV. $a'$ is the half of lattice constant $a$ (=6.1628(6)$\AA$) of the face-center-cubic unit cell. Green solid line indicates the momentum location of the measured bands in (b) and (c). (b), (c) Photoemission intensity plot along $\overline{\Gamma}$-$\overline{M}$ and corresponding 2D curvature intensity plot\cite{Aprecisemethod2011}, respectively.}
\end{figure}

ARPES measurements were performed to investigate the intrinsic electronic structures of YSb. As illustrated in Fig. 6(a), the topology of Fermi surfaces (FSs) is basically consistent with the previous ARPES\cite{Zeng2016,XHNiu2016} and calculation\cite{guo2016perfect,AkiraHasegawa1985} results on LaSb/LaBi, consisting of two hole pockets at the Brillouin zone (BZ) center and one elliptical electron pocket at BZ corner. Moreover, we observed some additional FSs around the $\overline{M}$ points, which could result from the band folding effect associated with lattice periodic potential of the termination layer on the (0 0 1) surface. The detailed band dispersions along $\overline{\Gamma}$-$\overline{M}$ are shown in Fig. 6(b) and (c), whose momentum location is indicated in Fig. 6(a). On moving from $\overline{\Gamma}$ to $\overline{M}$, the outer hole band gradually levels off and then curves upward, forming a hole band with a top at ~$-$0.75 eV at $\overline{M}$. Additionally, there is a parabolic electron band along $\overline{\Gamma}$-$\overline{M}$ with a bottom at ~$-$0.30 eV at $\overline{M}$, forming a band gap of ~0.45 eV. These band features are quite similar to that of LaSb\cite{Zeng2016}, which is demonstrated as a topologically trivial material without band anti-crossing along $\overline{\Gamma}$-$\overline{M}$. Combining with previous results and discussion on the SdH oscillation, we can conclude that YSb is also a topologically trivial material.

\begin{figure}[htbp]
\centering
\includegraphics[width=0.48\textwidth]{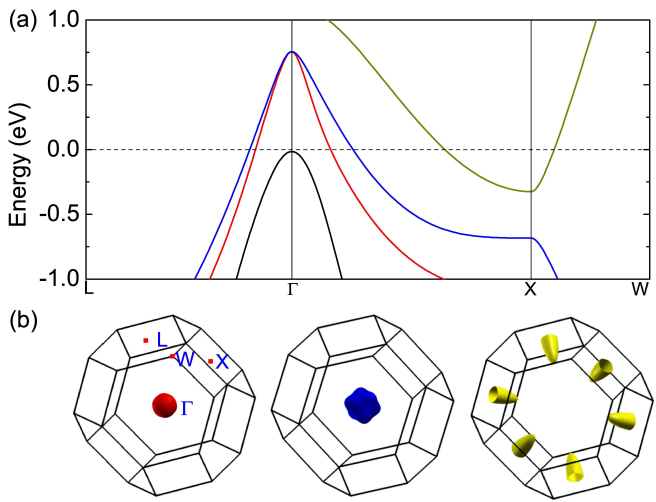}
\caption{(Color online) (a) Band structure along high-symmetry directions of the Brillouin zone and (b) FSs of YSb calculated with the MBJ potential and including the SOC effect.}
\end{figure}

The electronic structures of YSb have also been studied from first-principles calculations. As shown in Fig. 7(a), there are three bands crossing $E_F$ (there is also a band near the FS but slightly lower than the Fermi level), including two hole-type bands centered at $\Gamma$ point and one electron-type band around $X$ point with the ellipsoidal FS. For the two hole-type bands, one has a nearly spherical FS and another has a FS stretched in the $<100>$ directions. The gap between valence band and conduction band at $X$ point is about 0.354 eV. The band structure from first-principles calculation is consistent with the results of ARPES.

Now we turn to the discussion on the origin of the XMR in YSb. In Dirac semimetal Cd$_3$As$_2$, the XMR is attributed to the lifting of topological protection by the applied magnetic field\cite{liang2015ultrahigh}. The protection strongly suppresses backscattering in zero field and leads to a much longer transport lifetime than the quantum lifetime ($\tau_{tr}/\tau_Q\sim10^4$). In YSb, the transport life time is $\tau_{tr}^{\alpha}=\mu_em_{\alpha}^*/e=3.89\times10^{-12}s$, so the ratio of $\tau_{tr}^{\alpha}/\tau_Q^{\alpha}$ is about 57. This value is quite small compared with that in Cd$_3$As$_2$, indicating the lack of topological protection for the $\alpha$ band of YSb. This is also consistent with the topological trivial characteristic of YSb as described above. In the rare earth-based materials LaSb and LaBi, according to previous work\cite{sun2016large,kumar2016observation,guo2016perfect}, the perfect electron-hole compensation and high carrier mobilities naturally explain the XMR based on the two-band model. At the same time, results in another work\cite{tafti2016temperature} show that a combination of electron-hole compensation and the orbital texture on the electron band plays the key role in determining the magnitude of XMR. In YSb, electron-hole compensation is not perfect. However, the larger gap than LaSb/LaBi at $X$ point revealed by ARPES and first-principles calculations indicates that the orbital texture may not be suitable to explain the XMR in YSb. Therefore, it is suggested that the XMR in YSb still originates from electron-hole compensation and high mobility of carriers.

\section{Summary}

In summary, single crystals of YSb have been grown and the magneto-transport properties have been studied in detail. Field-induced metal-to-insulator-like transition and XMR are observed and the resistivity plateau emerges after the insulator-like regime. Moreover, at high magnetic field and low temperature, clear SdH oscillation appears in YSb. The FFT spectra reveals that there exist two major frequencies in the oscillation and the corresponding effective masses are extracted. The linear fitting in the Landau index plot gives a trivial Berry phase. Combining with the electronic structure revealed by ARPES experiments and first-principles calculations, we conclude that YSb is a topologically trivial material. The XMR in YSb can be attributed to electron-hole compensation and high mobility of carriers.

Note added: On preparing this paper for submission, we noticed one
similar work reported by Ghimire \emph{et al}. on the
magneto-transport properties of YSb, where similar results were
reported\cite{ghimire2016magnetotransport}. While our paper was
under revision, two more papers related got published with one
reporting on the magneto-transport
properties\cite{pavlosiuk2016giant} and another on ARPES study of
electronic structure in YSb where the topologically trivial nature
was revealed\cite{he2016distinct}, both of which cited this paper in
its first version. The difference of this paper from their work is
that the Landau level fan diagram is clearly presented and the Berry
phase extracted reveals a possible topologically trivial nature. The
value of $\tau_{tr}^{\alpha}/\tau_Q^{\alpha}$ is also discussed,
where the small value indicates the lack of topological protection.
All these suggest YSb is a topologically trivial metal, which is
further confirmed by ARPES and first-principles calculations.

\section{Acknowledgments}

We thank Prof. Z.-Y. Lu for his helpful discussions. This work is
supported by the National Natural Science Foundation of China
(No.11574391, No.11274381), the Fundamental Research Funds for the
Central Universities, and the Research Funds of Renmin University of
China (RUC) (No.14XNLQ07 and No.14XNLQ03). Computational resources
have been provided by the Physical Laboratory of High Performance
Computing at RUC. The Fermi surfaces were prepared with the XCRYSDEN
program\cite{kokalj2003computer}.

\bibliography{bibtex}
\end{document}